\documentclass[8pt,aps,twoside,nofootinbib]{revtex4}
\usepackage{graphicx}
\usepackage{amsmath,amssymb,amsthm, verbatim}
\usepackage{enumerate}
\usepackage{mathrsfs}
\usepackage{latexsym}
\usepackage{fancyhdr}
\newtheorem{prop}{Proposition}
\newtheorem{thm}[prop]{Theorem}
\newtheorem{cor}[prop]{Corollary}

\newcommand{\two}{2}
\newcommand{\defeq}{:=}

\newcommand{\lp}{l}
\newcommand{\vertex}{v}
\newcommand{\edge}{e}
\newcommand{\ext}{n}
\newcommand{\multiedges}{\rho}
\newcommand{\Qi}{Q_{i}}
\newcommand{\Qti}{{q_i}^{(\rho)}_{\ge 1}}
\newcommand{\Qtione}{{Q_i}^{(1)}_{\ge 1}}
\newcommand{\Qtialg}{{Q_i}^{(\rho)}_{\ge 1}}
\newcommand{\Qtifac}{{\hat{q_i}}^{(\rho)}_{\ge 1}}
\newcommand{\Qtifacalg}{{\hat{Q_i}}^{(\rho)}_{\ge 1}}
\newcommand{\Qcutalg}{\hat{Q}^{(\rho)}_{{v-1}_{\ge 1}}}
\newcommand{\copti}{{\Delta_i}_{\ge 1}}
\newcommand{\copt}{{\Delta}_{\ge 1}}
\newcommand{\ri}{r_i}
\newcommand{\Db}{\mathfrak{V}}
\newcommand{\Ds}{\Gamma}
\newcommand{\De}{\mathfrak{I}}
\newcommand{\Daux}{\mathfrak{B}}
\newcommand{\onevi}{B}
\newcommand{\graph}{S}
\newcommand{\B}{\mathscr{B}}
\newcommand{\BB}{\mathfrak{B}}
\newcommand{\mirror}{\tau}
\newcommand{\glue}{\Diamond}
\newcommand{\copba}{\bigtriangleup}
\newcommand{\copbai}{\bigtriangleup_i}
\newcommand{\copbaiplus}{\bigtriangleup_{i+1}}
\newcommand{\copbaj}{\bigtriangleup_j}

\newcommand{\C}{\mathbb{C}}
\DeclareMathOperator{\cou}{\epsilon}
\DeclareMathOperator{\cop}{\Delta}
\newcommand{\one}{1}
\DeclareMathOperator{\id}{id}
\newcommand{\Rop}{R}

\newcommand{\Gcompl}{G}
\newcommand{\Gfeyn}{G_F}
\newcommand{\Gconn}{G_c}
\newcommand{\Gopi}{G_{\text{1PI}}}

\newcommand{\compl}{\rho}
\newcommand{\conn}{\sigma}

\newcommand{\opi}{\tau}

\newcommand{\SV}{\mathsf{S}(V)}
\newcommand{\SVk}{\mathsf{S}^k(V)}
\newcommand{\SVi}{\mathsf{S}^i(V)}
\newcommand{\SVn}{\mathsf{S}^n(V)}
\newcommand{\SVni}{\mathsf{S}^{n-i}(V)}
\newcommand{\SVzero}{\mathsf{S}^0(V)}
\newcommand{\TV}{\mathsf{T}(\mathsf{S}(V))}

\newcommand{\fct}{\nu}
\newcommand{\Top}{T}
\newcommand{\exxt}{\mbox{ext}}
\newcommand{\innt}{\mbox{int}}
\newcommand{\xx}{\bar{s}}
\newcommand{\copb}{\vartriangle}

\newcommand{\tens}{\otimes}
\newcommand{\xd}{\mathrm{d}}
\newcommand{\Vertex}{\mathscr{V}}
\newcommand{\fctpi}{\nu_{{\tiny\mbox{1PI}}}}
\newcommand{\Tv}{\Sigma}
\newcommand{\uu}{u}
\newcommand{\RV}{\mathsf{R}(V)}
\newcommand{\RVk}{\mathsf{R}^k(V)}
\newcommand{\RVzero}{\mathsf{R}^0(V)}
\begin{document}
\title{Combinatorics of 1-particle irreducible $n$-point functions via coalgebra in quantum field theory}
\author{{\^A}ngela Mestre}\email{mestre@langite.impmc.jussieu.fr}

\affiliation{Institut de Min\'eralogie et de Physique des Milieux Condens\'es,
CNRS UMR 7590, Universit\'es Paris 6 et 7, IPGP, 140 rue de Lourmel,
75015 Paris, France.}
%\date{today}

%%%%%%%%%%%%%%%%%%%%%%%%%%%%%%%%%%%%%%%%%%%%%%%%%%%%%%%%%%%%%%%%%%%%%%%%%%%%%%%%%%%%%%%%%%%%%%%%%%%%%%%%%%%%%%%%%%%%%%%%%%%%%%%%%%%%%%%%%%%%%%%%%%%%%%%%%%%%%%%%%%%%%%%%%%%%%%%%%%%%%%%%%%%%%%%%%%%%%%%%%%%%%%%%%%%%%%%%%%%%%%%%%%%%%%%%%%%%%%%%%%%%%%%%%%%%%%%%%%%%%%%%%%%%%%%%%%%%%%%%%%%%%%%%%%%%%%%%%%%%%%%%%%%%%%%%%%%%%%%%%%%%%%%%%%%%%%%%%%%%%%%%%%%%%%%%%%%%%%%%%%%%%%%%%%%%
\begin{abstract}
We give a coalgebra structure on 1-vertex irreducible %(i.e., biconnected) 
graphs which is that of a cocommutative coassociative graded connected 
coalgebra. 
We generalize the coproduct to  the algebraic representation of graphs so as to
express a bare 
1-particle irreducible  $n$-point function in terms of its loop order contributions. 
The algebraic representation is so that graphs can be evaluated as Feynman graphs.
\end{abstract}
%%%%%%%%%%%%%%%%%%%%%%%%%%%%%%%%%%%%%%%%%%%%%%%%%%%%%%%%%%%%%%%%%%%%%%%%%%%%%%%%%%%%%%%%%%%%%%%%%%%%%%%%%%%%%%%%%%%%%%%%%%%%%%%%%%%%%%%%%%%%%%%%%%%%%%%%%%%%%%%%%%%%%%%%%%%%%%%%%%%%%%%%%%%%%%%%%%%%%%%%%%%%%%%%%%%%%%%%%%%%%%%%%%%%%%%%%%%%%%%%%%%%%%%%%%%%%%%%%%%%%%%%%%%%%%%%%%%%%%%%%%%%%%%%%%%%%%%%%%%%%%%%%%%%%%%%%%%%%%%%%%%%%%%%%%%%%%%%%%%%%%%%%%%%%%%%%%%%%%%%%%%%%%%%%%%%

\maketitle

\bigskip
%%%%%%%%%%%%%%%%%%%%%%%%%%%%%%%%%%%%%%%%%%%%%%%%%%%%%%%%%%%%%%%%%%%%%%%%%%%%%%%%%%%%%%%%%%%%%%%%%%%%%%%%%%%%%%%%%%%%%%%%%%%%%%%%%%%%%%%%%%%%%%%%%%%%%%%%%%%%%%%%%%%%%%%%%%%%%%%%%%%%%%%%%%%%%%%%%%%%%%%%%%%%%%%%%%%%%%%%%%%%%%%%%%%%%%%%%%%%%%%%%%%%%%%%%%%%%%%%%%%%%%%%%%%%%%%%%%%%%%%%%%%%%%%%%%%%%%%%%%%%%%%%%%%%%%%%%%%%%%%%%%%%%%%%%%%%%%%%%%%%%%%%%%%%%%%%%%%%%%%%%%%%%%%%%%%%
%\input{intro}
\section{Introduction}\label{sec:intro}
In perturbative quantum field theory the $n$-point  functions are given by weighted  sums over all Feynman graphs of a given type. Each Feynman graph represents an integral which is often divergent. This is due to the fact that products of Feynman propagators with coincident spacetime points are not defined. In particular, all divergences of a given Feynman graph reside in the 1-particle irreducible (1PI) subgraphs. Therefore, these are 
 essential for 
 renormalization 
for it is enough to renormalize the 1PI %Feynman
graphs.   
The systematic generation of all (bare) 1PI Feynman graphs is traditionally dealt with via the Legendre transform of the generating functional of connected Green functions  and iteration of the 1PI Dyson-Schwinger equations. 
Though this is the standard procedure, any method to straightforwardly manipulate 1PI Feynman graphs may actually be used. 
Here, we focus on the   algorithm to generate  all 1PI (i.e., 2-edge connected) graphs given in Ref. \onlinecite{Me:biconn}. This is so that larger 1PI graphs are produced from smaller ones  by increasing the number of their building blocks, the maximal 1-vertex irreducible (1VI) (i.e., biconnected)  subgraphs. The decomposition of 1PI  graphs in terms of their 1VI components is of interest to perturbative quantum field theory for the counterterm of a 1PI graph is given by the product of the counterterms of its 1VI components \cite{KS-F:critical}. 

The present  work is closely related to Ref. \onlinecite{MeOe:npoint,MeOe:loop} where recursion formulas to generate all   trees and all  
connected graphs are given. The underlying structure  is
an algebraic representation of graphs based on the    Hopf algebra structure of the  symmetric algebra on monomials of time-ordered field operators given in Ref. \onlinecite{BrOe:scalar, Br:meets}. 
This allowed to derive simple algebraic relations between complete, connected and 1PI $n$-point functions \cite{MeOe:npoint}, and to express a connected $n$-point function in terms of its loop order contributions \cite{MeOe:loop}.
Here, we extend the latter result
to 
1PI 
 $n$-point functions. 
To this end, we give a coalgebra structure on 1VI Feynman graphs which is that of a cocommutative coassociative graded connected
coalgebra. We generalize the coproduct to a map on the algebraic representation whose action is analogous to that of the coproduct. 
We use the truncated coproduct  \cite{MeOe:npoint} and the analog of the coproduct on 1VI graphs to derive algebraic expressions of  the recursion formulas to generate all  1VI and all
1PI graphs given in Ref. \onlinecite{Me:biconn}. 
 As in Ref. \onlinecite{MeOe:npoint,MeOe:loop}, the underlying algorithms are amenable to direct
implementation and  allow efficient calculations of graphs as well as their values as Feynman graphs.   Moreover, all results apply both to bosonic
and fermionic fields. Note, however, that no type of renormalization procedure is taken into
account. In this sense the computed $n$-point functions may
be considered as bare ones. 

This paper is organized as follows: Section \ref{sec:fey} recalls the definition of  Feynman graphs and their relation to the $n$-point functions.  Section~\ref{sec:field} reviews the Hopf algebra structure of the associative algebra on monomials of  time-ordered field operators given in Ref. \onlinecite{BrOe:scalar,Br:meets}.  Section~\ref{sec:algrep}  recalls the Hopf algebraic representation of graphs given in  Ref. \onlinecite{MeOe:npoint, MeOe:loop} and gives a non-associative graph multiplication.
Section \ref{sec:coalg} gives a coalgebra structure on 1VI Feynman graphs and generalizes the coproduct to the algebraic representation. 
Section \ref{sec:linear} gives the basic linear maps to construct 1VI or 1PI graphs.
Sections \ref{sec:1vi} and \ref{sec:1pi}  express the recursion formulas to generate 1VI and 1PI graphs given in Ref. \onlinecite{Me:biconn} 
 in a completely algebraic language, while Section \ref{sec:npoint} deals with their interpretation 
 in terms of 
 the 1PI $n$-point functions. 
%%%%%%%%%%%%%%%%%%%%%%%%%%%%%%%%%%%%%%%%%%%%%%%%%%%%%%%%%%%%%%%%%%%%%%%%%%%%%%%%%%%%%%%%%%%%%%%%%%%%%%%%%%%%%%%%%%%%%%%%%%%%%%%%%%%%%%%%%%%%%%%%%%%%%%%%%%%%%%%%%%%%%%%%%%%%%%%%%%%%%%%%%%%%%%%%%%%%%%%%%%%%%%%%%%%%%%%%%%%%%%%%%%%%%%%%%%%%%%%%%%%%%%%%%%%%%%%%%%%%%%%%%%%%%%%%%%%%%%%%%%%%%%%%%%%%%%%%%%%%%%%%%%%%%%%%%%%%%%%%%%%%%%%%%%%%%%%%%%%%%%%%%%%%%%%%%%%%%%%%%%%%%%%%%%%%

%\input{basicsfey}
\section{Basics}
\label{sec:basics}
In the following we are concerned with a generic perturbative
quantum field theory with fields $\phi(x)$ as operator valued distributions with adequate test functions. 
 Here, 
 $x$ represents a label that specifies spacetime coordinates, particle type, Minkowski indices, spin, colour, etc. The elementary field operators are local fields whose propagation and interactions may be described by local Hamiltonian or Lagrangian densities. For more information on these, we refer the reader to any standard textbook on quantum field theory such as Ref. \onlinecite{ItZu:qft}.   Note, however, that the essential property used in this paper is that they are (labeled) elements of
a vector space. Moreover, as in Ref. \onlinecite{MeOe:npoint,MeOe:loop}, for simplicity 
we limit
ourselves in the following exposition to a purely bosonic theory with a single scalar field. 
A more extensive discussion including fermionic fields is given in Section VI D of Ref. \onlinecite{MeOe:npoint}.
\subsection{Feynman graphs}\label{sec:fey}
We review the essentials about Feynman graphs that can be found in any textbook on quantum field theory such as Ref. \onlinecite{ItZu:qft}. 

\medskip

The following notation will be used in all the paper: by $\gamma$, we denote a (vertex) numbered graph while by $\bar{\gamma}$, we  denote the associated Feynman graph.

\bigskip
In perturbative quantum field theory, the calculation of physical quantities consists of evaluating integrals  represented by
Feynman graphs. These  are graphs that carry certain labels on vertices and
edges. The former correspond to the interaction terms of the
Lagrangian, while the latter may correspond to momenta and internal
indices (usually indicated by different styles of lines, e.g.,
straight for fermions, wiggly for bosons, etc).
Here, we assume that labels  are attached to the free ends of external edges, so that
internal labels shall be summed or integrated over.

\begin{figure}[t]
\begin{center}
\includegraphics[width=9cm]{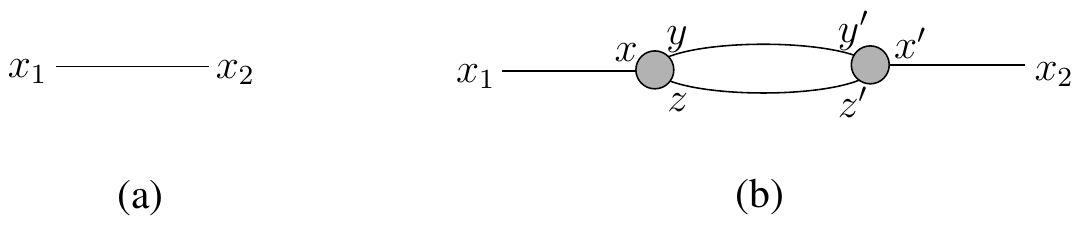}
\end{center}
\caption{Examples of Feynman graphs. (a) The Feynman propagator;
(b)
A contribution to a interaction of  the fields $\phi(x_1)$ and $\phi(x_2)$.}
\label{fig:feynman}
\end{figure}

The simplest Feynman graph is the graph  consisting of an external edge only, its two end points labeled  $x_1$ and $x_2$, respectively, and its value corresponding to the Feynman propagator $\Gfeyn(x_1,x_2)$, see Figure \ref{fig:feynman} (a).  The latter  is a Green's function for the free Klein-Gordon equation which is interpreted as  the probability amplitude that a particle changes position or some other attribute.   %The Feynman
%propagator is symmetric in its arguments $\Gfeyn(x_1,x_2)=\Gfeyn(x_2,x_1)$ as
%suggested by the corresponding symmetry of the graph.
The Feynman propagator is invertible under convolution. Its inverse $\Gfeyn^{-1}$  is determined by the following
equation: 
\begin{equation}
\int \xd y\, \Gfeyn(x,y)\Gfeyn^{-1}(y,z)=\delta(x,z)\,.
\label{eq:invfey}
\end{equation}

Besides changing spacetime positions or other properties, a particle can split into many other particles. The probability amplitude that this happens  is represented by (bare) interacting vertices. The value of a graph that consists of a vertex with external edges labeled $x_1,\dots,x_n$, is given by the $n$-point vertex function  $\Vertex^{(n)}(x_1,\dots,x_n)$:
\begin{equation}\label{eq:vertex}
\Vertex^{(n)}(x_1,\ldots,x_n)\defeq \int \xd y_1\ldots \xd y_n\, \Tv^{(n)}(y_1,\ldots,y_n) G_F(x_1,y_1)\ldots G_F(x_n,y_n)\,,
\end{equation}
where $\Tv^{(n)}$ denotes the truncated $n$-point vertex function.
In general, the value of a Feynman graph may  be computed as follows:
associate a label with each of the ends of internal edges and form the product over
a vertex function associated with each vertex and an inverse Feynman
propagator associated with each   internal edge. Finally, integrate over
all possible assignments of internal labels. For instance, the Feynman  graph given in Figure \ref{fig:feynman} (b) represents the probability amplitude that a particle in state $x_1$ goes to state $x$, splits into two particles in states $y$ and $z$, which go to states $y'$ and $z'$, respectively, and then fuse into a particle in state $x'$, which  goes to state $x_2$.  The value of this graph is given by
\begin{eqnarray*}
\lefteqn{\int dx dy dz dx' dy' dz'
G_F(x_1,x)\Tv^{(3)}(x,y,z)G_F(y,y')G_F(z,z')\Tv'^{(3)}(x',y',z')G_F(x',x_2)}\\
& & = \int dy dz dy' dz' \Vertex^{(3)}(x_1,y,z)G_F^{-1}(y,y')G_F^{-1}(z,z')\Vertex^{(3)}(x_2,y',z')\,.
\end{eqnarray*}

 A typical experiment consists of a setup of the initial particle configurations followed by a measurement of the final configurations. The theoretical prediction is expressed in terms of the Green's functions.  These are sums of the probability amplitudes associated with all possible ways in which the final state can be reached. For instance, the complete $n$-point function
$\Gcompl^{(n)}(x_1,\dots,x_n)$  is the vacuum expectation value of
the time-ordered product of $n$ field operators, i.e.,
\[
 \Gcompl^{(n)}(x_1,\dots,x_n)
 =\langle 0 | T \phi(x_1)\cdots \phi(x_n) | 0\rangle ,
\]
where $T$ denotes the {\em time-ordering} operator (see Ref. \onlinecite{ItZu:qft} for instance).  $T$
arranges the fields in decreasing time, reading from left to right, so that $\phi$ commutes with itself at equal times.

In terms of Feynman graphs the  various types of $n$-point functions
 are given by 
  the sum over the values of all
Feynman graphs (of  the same kind) with exactly $n$ external edges labeled  $x_1,\dots,x_n$. Each graph is associated with a scalar factor, called \emph{weight} which corresponds to the inverse of its symmetry factor.  For instance, denote by $\mathfrak{G}^n$ the set of all Feynman graphs of the given
theory with $n$ external edges. Given an arbitrary Feynman graph
$\bar{\gamma}\in\mathfrak{G}^n$,  denote by $\bar{\gamma}(x_1,\dots,x_n)$ its value for
the given labelings. In this context, the complete $n$-point function yields
\begin{equation}
 \Gcompl^{(n)}(x_1,\dots,x_n)=\sum_{\bar{\gamma}\in\mathfrak{G}^n}
 w_{\bar{\gamma}}\bar{\gamma}(x_1,\dots,x_n) ,
\label{eq:npfeyn}
\end{equation}
where $w_{\bar{\gamma}}$ denotes the weight of the graph $\bar{\gamma}$.

In the following, we will consider the restricted classes of Feynman graphs which are $1$-vertex or $1$-particle irreducible. We recall here the definitions (see Ref. \onlinecite{ItZu:qft} (pages 289 and 393)): 
A connected graph is said to be \emph{$1$-vertex irreducible (1VI)} (resp. \emph{$1$-particle irreducible (1PI)}) if and only if it remains connected after erasing any vertex (together with assigned edges) (resp. any internal edge). We consider that a single vertex is both 1VI and 1PI by convention. Clearly, all 1VI graphs but the $\two$-vertex tree are 1PI. 
Furthermore, given a connected graph, we call \emph{articulation} vertex to any vertex whose removal disconnects the graph. In this context, a 1VI graph is one without articulation vertices.
%%%%%%%%%%%%%%%%%%%%%%%%%%%%%%%%%%%%%%%%%%%%%%%%%%%%%%%%%%%%%%%%%%%%%%%%%%%%%%%%%%%%%%%%%%%%%%%%%%%%%%%%%%%%%%%%%%%%%%%%%%%%%%%%%%%%%%%%%%%%%%%%%%%%%%%%%%%%%%%%%%%%%%%%%%%%%%%%%%%%%%%%%%%%%%%%%%%%%%%%%%%%%%%%%%%%%%%%%%%%%%%%%%%%%%%%%%%%%%%%%%%%%%%%%%%%%%%%%%%%%%%%%%%%%%%%%%%%%%%%%%%%%%%%%%%%%%%%%%%%%%%%%%%%%%%%%%%%%%%%%%%%%%%%%%%%%%%%%%%%%%%%%%%%%%%%%%%%%%%%%%%%%%%%%%%%
%\input{basicsfield}
\subsection{The field operator algebra as a Hopf algebra}\label{sec:field}
We briefly recall the  Hopf algebra structure of the time-ordered field operator algebra given in  Ref. \onlinecite{BrOe:scalar, Br:meets}   (see also Ref. \onlinecite{BFFO:twist}).
Let $V$ denote the $\C$-vector space of finite linear combinations of elementary field
operators $\phi(x)$. Let $\SV$ denote  the free commutative unital associative algebra generated by all  time-ordered products of field operators.\footnote{For convenience, in $\SV$, we identify the time-ordered product with the free commutative product.}
Let $\SVk$ denote the vector space of
$k$-fold time-ordered products of field operators. Then, 
$\SV$ is the direct sum
of the spaces $\SVk$, i.e.,
$\SV=\bigoplus_{k=0}^\infty \SVk$, where $\SVzero\defeq\C1$.
This is a 
Hopf algebra \cite{Eisenbud, Kas, Loday} equipped with
\begin{enumerate}[(a)]
 \item a  
coproduct
$\cop:\SV\to \SV\tens \SV$ 
defined on $\C$ and $V$ by
$\cop(\one) \defeq \one\tens\one$,
$\cop(\phi(x)) \defeq  \phi(x)\tens \one + \one\tens\phi(x)$,
and extended to the whole of $\SV$ by compatibility with the product;
\item a counit
$\cou:\SV\to\C;\one\mapsto 1,\phi(x_1)\cdots\phi(x_n)\mapsto 0$ for $ n>0$;
\item an antipode  $S:\SV\to\SV;\one\mapsto \one,\phi(x_1)\dots\phi(x_n)\mapsto (-1)^n \phi(x_1)\dots\phi(x_n).$
\end{enumerate}
Furthermore, we define the   $k$-fold application of the coproduct,  $\cop^k:\SV\to
\SV^{\tens k+1}$, as follows: $\cop^0\defeq\id$ and
$\cop^{k+1}\defeq(\cop\tens\id^{\tens k})\circ\cop^k$. The latter
equation
can be written in $k+1$ different ways (corresponding to applying the coproduct to each of the $k+1$ tensor factors) which are all
equivalent due to the coassociativity of the coproduct.

The various types of $n$-point functions are vacuum expectation values of $n$ time-ordered field operators.
They correspond to the sum over the values of all Feynman graphs of a given type.
By $\Gcompl^{(n)}$, $\Gconn^{(n)}$ and  $\Gopi^{(n)}$, 
 we  denote complete, connected and  1PI 
$n$-point functions, respectively. Also, by $\Vertex^{(n)}_{\text{1PI}}$, we denote the 1PI $n$-point vertex functions. 
The ensemble of
time-ordered
$n$-point functions of a given type determine  maps
$\SV\to \C$  \cite{BFFO:twist}:
\begin{align*}
\compl(\phi(x_1)\cdots\phi(x_n)) & \defeq \Gcompl^{(n)}(x_1,\dots,x_n)\,,\\
\conn(\phi(x_1)\cdots\phi(x_n)) & \defeq \Gconn^{(n)}(x_1,\dots,x_n)\,,\\
\opi(\phi(x_1)\cdots\phi(x_n)) & \defeq \Gopi^{(n)}(x_1,\dots,x_n)\,,\\
\fctpi(\phi(x_1)\cdots\phi(x_n)) & \defeq \Vertex_{\text{1PI}}^{(n)}(x_1,\dots,x_n)\,.
\end{align*}
The assumption that
all 1-point functions vanish means that
$\compl(\phi(x))=\conn(\phi(x))=\opi(\phi(x))=\fctpi(\phi(x))=0$. 
Moreover,
the  0-point functions read as $\compl(\one)=1$,
$\conn(\one)=\opi(\one)=\fctpi(\one)=0$. 
%%%%%%%%%%%%%%%%%%%%%%%%%%%%%%%%%%%%%%%%%%%%%%%%%%%%%%%%%%%%%%%%%%%%%%%%%%%%%%%%%%%%%%%%%%%%%%%%%%%%%%%%%%%%%%%%%%%%%%%%%%%%%%%%%%%%%%%%%%%%%%%%%%%%%%%%%%%%%%%%%%%%%%%%%%%%%%%%%%%%%%%%%%%%%%%%%%%%%%%%%%%%%%%%%%%%%%%%%%%%%%%%%%%%%%%%%%%%%%%%%%%%%%%%%%%%%%%%%%%%%%%%%%%%%%%%%%%%%%%%%%%%%%%%%%%%%%%%%%%%%%%%%%%%%%%%%%%%%%%%%%%%%%%%%%%%%%%%%%%%%%%%%%%%%%%%%%%%%%%%%%%%%%%%%%%%
%\input{algrep}

\section{A Hopf algebraic representation of graphs}
\label{sec:algrep}
We  review the correspondence between graphs and certain elements of $\SV^{\tens\vertex}$ given in Ref. \onlinecite{MeOe:npoint, MeOe:loop}.
We consider the tensor algebra $\TV$ generated by the %graded 
vector space $\SV$. We recall the definition of tensor concatenation and give a non-associative multiplication in   $\TV$. 

\subsection{Correspondence between graphs and elements of $\SV^{\tens\vertex}$}\label{sec:corresp}
We use the Hopf algebraic representation of graphs given in  Ref. \onlinecite{MeOe:npoint, MeOe:loop} to represent 
graphs 
by tensors whose indices correspond to the vertex numbers.  

\bigskip

For all integers $\vertex\ge 1$, we  define the  elements $\Rop_{i,j}, \Rop_{i,i}\in\SV^{\tens\vertex}$ following  Ref. \onlinecite{MeOe:npoint,MeOe:loop}:\footnote{The integrations in equations (\ref{eq:Rij}) and (\ref{eq:Rii}) are part of the definitions of $\Rop_{i,j}$ and  $\Rop_{i,i}$ to simplify the notation in the following. Indeed, only the integrands are actually elements of $\SV^{\tens\vertex}$. To be precise, the integrations over all internal variables would be included later on, on the rhs of the formula of Corollary \ref{cor:loopexpansion}.}
\begin{equation}\label{eq:Rij}
\Rop_{i,j} := \int
\xd x\,\xd y\,G_F^{-1}(x,y)\,(\one^{\tens i-1}\tens \phi(x)
\tens\one^{\tens j-i-1}
\tens\phi(y)\tens\one^{\tens v-j}) ,
\end{equation}
where  the field operators $\phi(x)$ and $\phi(y)$ are
inserted at the
positions $i$ and $j$, respectively, with $i\not=j$. In the equation above  $G_F^{-1}(x,y)$ denotes the inverse Feynman propagator given by formula (\ref{eq:invfey}). For $i=j$ the definition is
\begin{equation}\label{eq:Rii}
\Rop_{i,i} := \int
\xd x\,\xd y\,G_F^{-1}(x,y)\,(\one^{\tens i-1}\tens \phi(x)
\phi(y)\tens\one^{\tens v-i})\,.
\end{equation}
For simplicity our notation does not distinguish between elements, say, $\Rop_{i,j},\Rop_{i,i}\in\SV^{\tens\vertex}$ and  $\Rop_{i,j},\Rop_{i,i}\in\SV^{\tens\vertex'}$ with $\vertex\neq\vertex'$. This convention will often be used in the rest of the paper for all elements of the algebraic representation. Therefore, we shall specify the set containing the given elements whenever necessary.

\smallskip

We  proceed to the correspondence
 between graphs on $\vertex$
vertices and elements of $\SV^{\tens \vertex}$ established in the aforesaid papers. For all $1\le i,j\le \vertex$ with $i\neq j$:
\begin{itemize}
\item
a  tensor factor in the $i$th position  corresponds to a
vertex
numbered  $i$; 
\item
 a product
$\phi(x_1)\cdots\phi(x_n)$ in the $i$th tensor factor corresponds to a vertex numbered $i$ which is assigned with $n$ external edges  whose end points are labeled $x_1,\dots,x_n$;
\item
$\Rop_{i,j}\in\SV^{\tens\vertex}$ corresponds to an
internal edge connecting
the vertices $i$ and $j$;
\item
$\Rop_{i,i}\in\SV^{\tens\vertex}$ 
corresponds to  a self-loop of
the vertex $i$.
\end{itemize}
\begin{figure}[t]
\begin{center}
\includegraphics[width=9cm]{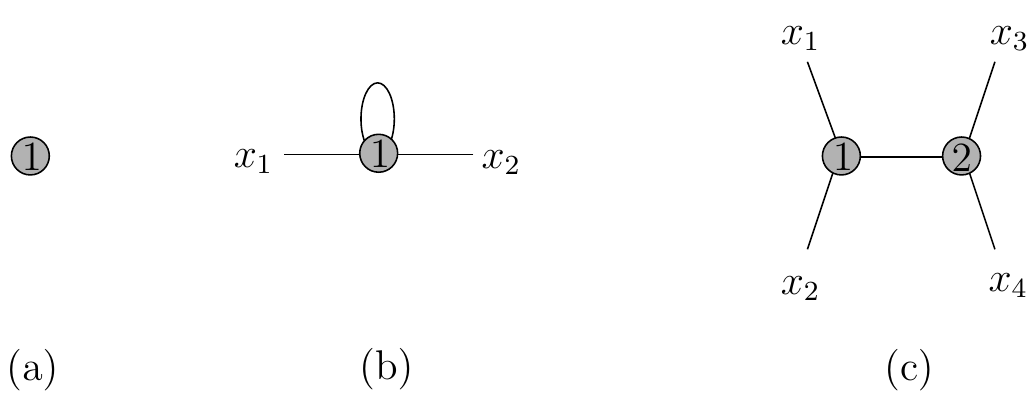}
\end{center}
\caption{(a) An isolated vertex represented by $\one\in\SV$; (b) A vertex with a self-loop and two external edges represented by $\Rop_{1,1}\cdot(\phi(x_1)\phi(x_\two))\in\SV$; (c) A tree graph on two vertices and two external edges represented by $\Rop_{1,\two}\cdot(\phi(x_1)\phi(x_\two)\tens\phi(x_3)\phi(x_4))\in\SV^{\tens\two}$.}
\label{fig:3graphs}
\end{figure}
Figure~\ref{fig:3graphs} shows some examples of this correspondence. 

\smallskip
 
Now, let us recall the definition of  the componentwise product $\cdot:\SV^{\tens\vertex}\times \SV^{\tens\vertex}\to\SV^{\tens\vertex};(\uu_1\tens\ldots\tens \uu_{\vertex},\uu'_1\tens\ldots\tens \uu'_{\vertex})\mapsto \uu_1\uu'_1\tens\ldots\tens \uu_{\vertex} \uu'_{\vertex}$,
where $\uu_i,\uu'_j$ denote monomials on the elementary field operators for all $1\le i,j\le\vertex$. 
Clearly, combining several internal  and
external edges by multiplying their expressions in
$\SV^{\tens \vertex}$ allows to build arbitrary graphs on $\vertex$
vertices. Applying a given type of vertex functions 
$\fct^{\tens \vertex}$ to the resulting expression 
yields the
value of the associated Feynman graph in the sense of Section \ref{sec:fey}. 
Recall that the vertex functions carry a Feynman propagator for all assigned ends of edges (see equation (\ref{eq:vertex})). Hence,  
 the inverse
Feynman propagator $\Gfeyn^{-1}$ in the definition of the elements $\Rop_{i,j},\Rop_{i,i}$ which represent internal edges,  cancels one superfluous
Feynman propagator. 

\begin{figure}[h]
\begin{center}
\includegraphics[width=5cm, height=3cm]{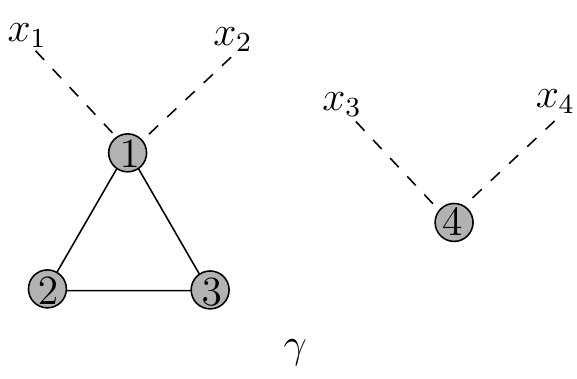}
\end{center}
\caption{A disconnected graph represented by the tensor $\graph_{1,\ldots,4}^\gamma=\innt^\gamma_{1,\ldots,4}\cdot\exxt^\gamma_{1,\ldots,4}\in\SV^{\tens 4}$.}
\label{fig:gamma}
\end{figure}

\begin{figure}[h]
\begin{center}
\includegraphics[width=9cm, height=2.2cm]{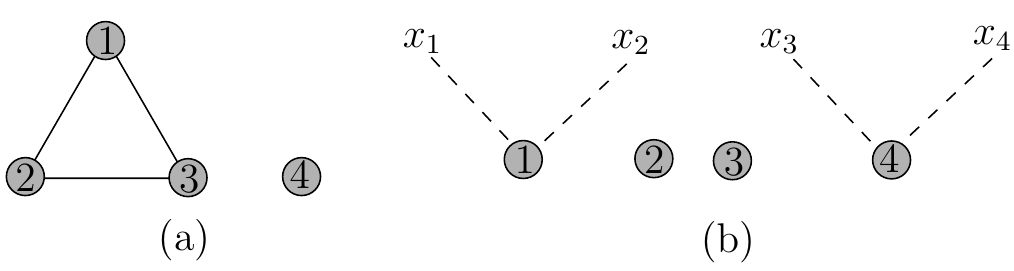}
\end{center}
\caption{(a) The internal edge tensor associated with the graph $\gamma$: $\innt^\gamma_{1,\ldots,4}=\Rop_{1,\two}\cdot\Rop_{1,3}\cdot\Rop_{\two,3}\in\SV^{\tens 4}$; (b) The external edge tensor associated with the graph $\gamma$: 
$\exxt^\gamma_{1,\ldots,4}=\phi(x_1)\phi(x_\two)\tens1\tens1\tens\phi(x_3)\phi(x_4)\in\SV^{\tens 4}$.} 
\label{fig:gammad}
\end{figure}

For instance, let $\gamma$ denote an arbitrary graph on $\vertex$ vertices and $\edge$ internal edges. For all $1\le k\le\edge$, let $\Rop_{i_k,j_k}$ represent an internal edge of $\gamma$ with $1\leq i_k,j_k\leq\vertex$. Also, for all $1\leq i\leq  \vertex$, let the vertex $i$ be assigned with the following $n_i\geq 0$ external edges $\phi(x_{1,i}),\ldots,\phi(x_{n_i,i})$ with $\phi(x_{0,i})\defeq1$.  In this context,  we define the \emph{internal edge tensor} associated with $\gamma$, $\innt_{1,\ldots,\vertex}^\gamma\in\SV^{\tens \vertex}$, as follows:
\begin{equation}
\innt_{1,\ldots,\vertex}^\gamma\defeq\prod_{k=1}^\edge \Rop_{i_k,j_k}\,.
\label{eq:int}
\end{equation} 
 Accordingly,  we define the  \emph{external edge tensor} associated with $\gamma$, $\exxt_{1,\ldots,\vertex}^\gamma\in\SV^{\tens \vertex}$,  by inserting  the external edges of the vertex $i$ in the $i$th tensor factor, for all $1\leq i\leq \vertex$. That is,
\begin{equation}
\exxt_{1,\ldots,\vertex}^\gamma\defeq\phi(x_{1,1})\ldots\phi(x_{n_1,1})\tens\ldots\tens\phi(x_{1,\vertex})\ldots\phi(x_{n_\vertex,\vertex})\,,\label{eq:ext}
\end{equation}
where the operator labels are all distinct. Clearly, both $\innt_{1,\ldots,\vertex}^\gamma$ and $\exxt_{1,\ldots,\vertex}^\gamma$ are elements of $\SV^{\tens\vertex}$ which represent graphs themselves. 
In this context, the algebraic representation is so that the graph $\gamma$ is uniquely represented by a tensor  $\graph^\gamma_{1,\ldots,\vertex}\in\SV^{\tens\vertex}$ given by the  (componentwise) product of  $\innt_{1,\ldots,\vertex}^\gamma$ by $\exxt_{1,\ldots,\vertex}^\gamma$.
That is,
\begin{equation}\graph^\gamma_{1,\ldots,\vertex}\defeq\innt_{1,\ldots,\vertex}^\gamma\cdot\exxt_{1,\ldots,\vertex}^\gamma\,.
\label{eq:tensgamma}
\end{equation} 
For instance, Figure \ref{fig:gamma} shows a disconnected graph, while Figure \ref{fig:gammad} %(a) and \ref{fig:gammad} (b) 
shows the associated internal and external edge tensors. 


Moreover,
the ordering of the tensor factors of $\SV^{\tens \vertex}$
determines the numbering of the vertices of a graph.   Note, however, that when applying a given type of vertex functions 
$\fct^{\tens \vertex}$, all mutually isomorphic graphs contribute to the value of the same Feynman graph. In this context, let $\C\mathfrak{G}_\vertex$ denote the $\C$-vector space on the set  $\mathfrak{G}_\vertex$ of all Feynman 
graphs (i.e., unnumbered graphs) on $\vertex$ vertices. The subset of $\SV^{\tens\vertex}$ whose elements are so that the labels of the underlying elementary field operators are all distinct, may be clearly embedded in   $\mathfrak{G}_\vertex$ via the correspondence $\gamma\mapsto\bar{\gamma}$, where $\bar{\gamma}$ denotes the Feynman graph associated with $\gamma$. We will not point this out explicitly in the following.

\begin{figure}[h]
\begin{center}
\includegraphics[width=7.5cm, height=3cm]{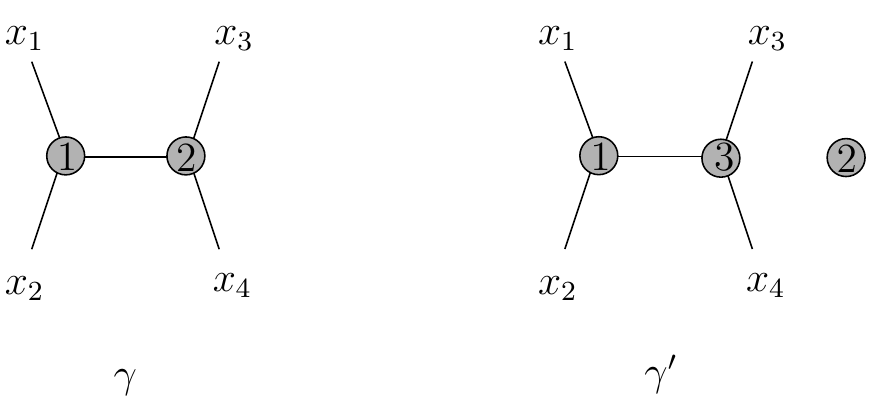}
\end{center}
\caption{The graph $\gamma$ is represented by the tensor $\graph^{\gamma}_{1,\two}\in\SV^{\tens\two}$; The graph $\gamma'$ is represented by the tensor $\graph^{\gamma}_{1,3}\in\SV^{\tens 3}$ associated with $\gamma$ and the bijection $\sigma:\{1,\two\}\to\{1,3\};1\mapsto 1,\two\mapsto 3$.} 
\label{fig:sigma}
\end{figure}

\medskip
Furthermore, let $\vertex'\geq\vertex$ denote an integer.  Let $X\subseteq\{1,\ldots,\vertex'\}$ be a set of cardinality $\vertex$. Let $\sigma:\{1,\ldots,\vertex\}\to X$ be a bijection. For all $\vertex'\geq\vertex$, we define the following elements of $\SV^{\tens\vertex'}$:
\begin{equation}
\graph_{\sigma(1),\ldots,\sigma(\vertex)}^\gamma\defeq\innt^\gamma_{\sigma(1),\ldots,\sigma(\vertex)}\cdot\exxt^\gamma_{\sigma(1),\ldots,\sigma(\vertex)}\,,\label{eq:sigma}
\end{equation}
where 
\begin{equation}
\innt^\gamma_{\sigma(1),\ldots,\sigma(\vertex)}\defeq\prod_{k=1}^\edge \Rop_{\sigma(i_k),\sigma(j_k)} ,
\label{eq:intsigma}
\end{equation}
\begin{equation}
\exxt^\gamma_{\sigma(1),\ldots,\sigma(\vertex)}\defeq\prod_{i=1}^{\vertex'}\one^{\tens \sigma(i)-1}\tens \phi(x_{1,i})\ldots\phi(x_{n_i,i})\tens
\one^{\tens v-\sigma(i)}\,.\label{eq:extsigma}
\end{equation}
In terms of graphs, the tensor $\graph_{\sigma(1),\ldots,\sigma(\vertex)}^\gamma$ represents a disconnected graph, say, $\gamma'$, on $\vertex'$ vertices consisting of a graph isomorphic to $\gamma$ whose vertices take numbers from the set $X$ and $\vertex'-\vertex$ isolated vertices taking numbers from the set $\{1,\ldots,\vertex'\}\backslash X$. Figure \ref{fig:sigma} shows an example. The interpretation of the tensors $\innt^\gamma_{\sigma(1),\ldots,\sigma(\vertex)}$ and $\exxt^\gamma_{\sigma(1),\ldots,\sigma(\vertex)}$ is analogous.
%%%%%%%%%%%%%%%%%%%%%%%%%%%%%%%%%%%%%%%%%%%%%%%%%%%%%%%%%%%%%%%%%%%%%%%%%%%%%%%%%%%%%%%%%%%%%%%%%%%%%%%%%%%%%%%%%%%%%%%%%%%%%%%%%%%%%%%%%%%%%%%%%%%%%%%%%%%%%%%%%%%%%%%%%%%%%%%%%%%%%%%%%%%%%%%%%%%%%%%%%%%%%%%%%%%%%%%%%%%%%%%%%%%%%%%%%%%%%%%%%%%%%%%%%%%%%%%%%%%%%%%%%%%%%%%%%%%%%%%%%%%%%%%%%%%%%%%%%%%%%%%%%%%%%%%%%%%%%%%%%%%%%%%%%%%%%%%%%%%%%%%%%%%%%%%%%%%%%%%%%%%%%%%%%%%%
%\input{tens}
\subsection{Tensor algebra}\label{sec:tens}
We consider the tensor algebra $\TV$ generated by the graded 
vector space $\SV$.  We use the concatenation of tensors  to define a non-associative product in $\TV$.  This may be interpreted in terms of graphs as the operation of gluing two graphs at a 
vertex.

\bigskip

Consider the tensor algebra on the graded vector space $\SV$: $\TV\defeq\bigoplus_{k=1}^{\infty}\SV^{\tens k}$. In $\TV$  the multiplication $\bullet:\TV\times\TV\to\TV$ is defined by  concatenation of tensors (see Ref. \onlinecite{k-theory} for instance):
\begin{equation}
(\uu_1\tens\ldots\tens \uu_\vertex)\bullet(\uu'_1\tens\ldots\tens \uu'_{\vertex'})\defeq
\uu_1\tens\ldots\tens \uu_\vertex\tens \uu'_1\tens\ldots\tens \uu'_{\vertex'}\,,\label{eq:conc}
\end{equation}
where $\uu_i\,,\uu'_j$ denote monomials on the elementary field operators for all $1\leq i\leq\vertex$, $1\leq j\leq \vertex'$.
We proceed to generalize the definition of the multiplication $\bullet$ to any two positions of the tensor factors. Let $\mirror:\SV^{\tens\two}\to\SV^{\tens\two};\uu_1\tens \uu_\two\mapsto \uu_\two\tens \uu_1$.  Moreover, define $\mirror_k\defeq
\id^{\tens{k-1}}\tens\mirror\tens\id^{\tens{\vertex-k-1}}:\SV^{\tens\vertex}\to \SV^{\tens\vertex}$ for all $1\leq k\leq\vertex-1$ . In this context, for all $1\leq i\leq\vertex$, $1\leq j\leq \vertex'$, define $\bullet_{i,j}:\SV^{\tens \vertex}\times \SV^{\tens \vertex'}\to \SV^{\tens \vertex+\vertex'}$ %: \TV\times\TV\to\TV$ 
by the following equation:
\begin{eqnarray}\nonumber
%\lefteqn{\bullet_{i,j}:\SV^{\tens \vertex}\times \SV^{\tens \vertex'}\to \SV^{\tens \vertex+\vertex'}}
 \lefteqn{(\uu_1\tens\ldots\tens \uu_\vertex)\bullet_{i,j}(\uu'_1\tens\ldots\tens \uu'_{\vertex'})\defeq}\\
& & (\mirror_{\vertex-1}\circ\ldots\circ\mirror_{i})(\uu_1\tens\ldots\tens \uu_\vertex)\tens(\mirror_{1}\circ\ldots\circ\mirror_{j-1})(\uu'_1\tens\ldots\tens \uu'_{\vertex'})\\\label{eq:concijI}
& & \quad= \uu_1\tens\ldots\tens\hat{\uu_i}\tens\ldots\tens \uu_\vertex\tens \uu_i\tens \uu'_j\tens \uu'_1\tens\ldots\tens\hat{\uu'_j}\tens\ldots\tens \uu'_{\vertex'}\,,\label{eq:concijII}
\end{eqnarray}
where $\hat{\uu_i}$ (resp. $\hat{\uu'_j}$) means that $\uu_i$ (resp. $\uu'_j$) is excluded from the sequence.
In terms of the tensors $\graph^\gamma_{1,\ldots,\vertex}\in\SV^{\tens\vertex}$ and $\graph^{\gamma'}_{1,\ldots,\vertex'}\in\SV^{\tens\vertex'}$ the equation above yields
\begin{equation}
\graph^\gamma_{1,\ldots,\vertex}\bullet_{i,j}\graph^{\gamma'}_{1,\ldots,\vertex'}\defeq\graph_{\sigma(1),\ldots,\sigma(\vertex)}^\gamma\cdot\graph_{\sigma'(1),\ldots,\sigma'(\vertex')}^{\gamma'}\,,\label{eq:conctens}
\end{equation}
where $\graph_{\sigma(1),\ldots,\sigma(\vertex)}^\gamma,\graph_{\sigma'(1),\ldots,\sigma'(\vertex')}^{\gamma'}\in\SV^{\tens(\vertex+\vertex')}$, $\sigma(k)=k$ if $1\leq k<i$, $\sigma(i)=\vertex$, $\sigma(k)=k-1$ if $i<k\leq\vertex$ and $\sigma'(k)=k+\vertex+1$ if $1\leq k<j$, $\sigma'(j)=\vertex+1$, $\sigma'(k)=k+\vertex$ if $j<k\leq\vertex'$. Clearly, $\graph_{1,\ldots,\vertex}^\gamma\bullet_{i,j}\graph_{1,\ldots,\vertex'}^{\gamma'}$ corresponds to a disconnected graph.
\begin{figure}[t]
\begin{center}
\includegraphics[width=8cm, height=2.7cm]{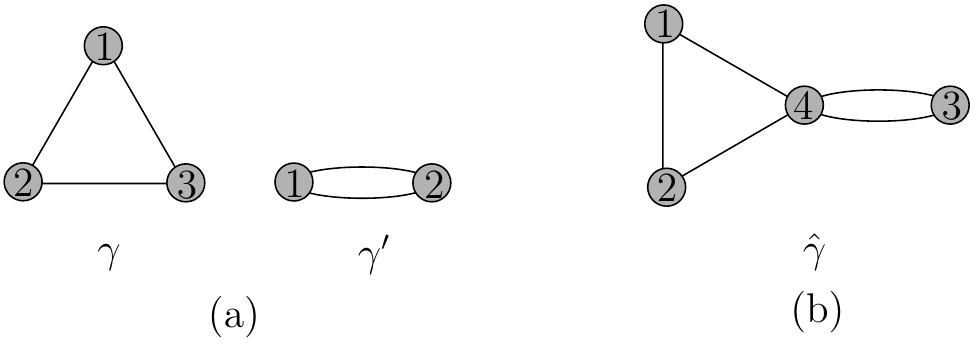}
\end{center}
\caption{(a) The graphs represented by the tensors $\graph_{1,\two,3}^\gamma\in\SV^{\tens 3}$ and $\graph_{1,\two}^{\gamma'}\in\SV^{\tens \two}$; (b) The graph represented by the tensor $\graph^{\hat{\gamma}}_{1,\two,3,4}=\graph_{1,\two,3}^\gamma\glue_{3,\two}\graph_{1,\two}^{\gamma'}\in\SV^{\tens 4}$.} 
\label{fig:glue}
\end{figure}
\smallskip
Now, let $\cdot_i\defeq
\id^{\tens{i-1}}\tens\cdot\tens\id^{\tens{\vertex-i-1}}:\SV^{\tens\vertex}\to\SV^{\tens(\vertex-1)}$. In $\TV$,  for all $1\leq i\leq\vertex$, $1\leq j\leq \vertex'$, define the following 
non-associative and non-commutative multiplication:  
\begin{equation}
\glue_{i,j}\defeq\mirror_{\vertex+\vertex'-2}\circ\ldots\circ\mirror_{\vertex}\circ\cdot_\vertex\circ\bullet_{i,j}:\SV^{\tens\vertex}\times\SV^{\tens\vertex'}\to\SV^{\tens(\vertex+\vertex'-1)}
\,.\\\label{eq:glue}
\end{equation}
The tensor $\graph^\gamma_{1,\ldots,\vertex}\glue_{i,j}\graph^{\gamma'}_{1,\ldots,\vertex'}$ represents the graph on $\vertex+\vertex'-1$ vertices, say, $\hat{\gamma}$ obtained by gluing the vertex $i$ of $\gamma$ to the vertex $j$ of $\gamma'$.
Note that the subgraphs $\gamma$ and $\gamma'$ of the graph $\hat{\gamma}$ share only the vertex $\vertex+\vertex'-1$  (together with assigned external edges or self-loops) and no internal edges (distinct from self-loops). The vertex $\vertex+\vertex'-1$  is clearly an articulation vertex of the graph $\hat{\gamma}$. Figure \ref{fig:glue} shows an example. 
%%%%%%%%%%%%%%%%%%%%%%%%%%%%%%%%%%%%%%%%%%%%%%%%%%%%%%%%%%%%%%%%%%%%%%%%%%%%%%%%%%%%%%%%%%%%%%%%%%%%%%%%%%%%%%%%%%%%%%%%%%%%%%%%%%%%%%%%%%%%%%%%%%%%%%%%%%%%%%%%%%%%%%%%%%%%%%%%%%%%%%%%%%%%%%%%%%%%%%%%%%%%%%%%%%%%%%%%%%%%%%%%%%%%%%%%%%%%%%%%%%%%%%%%%%%%%%%%%%%%%%%%%%%%%%%%%%%%%%%%%%%%%%%%%%%%%%%%%%%%%%%%%%%%%%%%%%%%%%%%%%%%%%%%%%%%%%%%%%%%%%%%%%%%%%%%%%%%%%%%%%%%%%%%%%%%
%\input{coalg}
\section{A coalgebra structure on 1VI graphs}\label{sec:coalg}
We give a coalgebra structure on 1VI Feynman graphs which is that of a cocommutative coassociative graded connected
coalgebra.
We generalize the coproduct to the algebraic representation.
We emphasize that in the present section only graphs  with no external edges nor self-loops are considered. Hence, in all that follows  by \emph{graphs} we mean graphs without  external edges nor self-loops. 

\bigskip

Let $\BB_{\lp,\vertex}$ denote the set of all 1VI Feynman
graphs 
on $\lp$ loops and  $\vertex$ vertices. 
Let $\C\BB_{\lp,\vertex}$ denote the $\C$-vector space with the set $\BB_{\lp,\vertex}$ as basis.
Let $\C\BB\defeq\bigoplus_{\vertex=1,\lp=0}^{\infty}\C\BB_{\lp,\vertex}$. 
We  define the following coalgebra structure on the graded vector space $\C\BB$:

\begin{itemize} 
\item the coproduct $\copb:\C\BB\to\C\BB\otimes\C\BB$ is defined 
as follows:
\begin{eqnarray}
\copb(\xx)&\defeq&\xx\tens\xx\,;\\
\copb(\bar{\gamma})&\defeq&\xx\otimes\bar{\gamma}+\bar{\gamma}\otimes\xx \quad\mbox{if}\quad\bar{\gamma}\neq\xx\,,\label{eq:cop1vibasic}
\end{eqnarray} 
where $\xx$ and $\bar{\gamma}$  denote a single vertex and an (unnumbered)  1VI  graph  on $\vertex>1$ vertices, respectively.  The coproduct $\copb$ is clearly coassociative and cocommutative;

\item
the \emph{counit}
$\cou:\C\BB\to\C$ is defined by 
\begin{eqnarray}
\cou(\xx)&\defeq&1\,;\\
\cou(\bar{\gamma})&\defeq&0\quad\mbox{if}\quad\bar{\gamma}\neq\xx\,.\label{eq:copbbasic}
\end{eqnarray}
\end{itemize}
\smallskip

We grade the coalgebra $\C\BB$ as follows: if $\bar{\gamma}\in\BB_{\lp,\vertex}$ then $\bar{\gamma}$ has degree $\vertex-1$.
$\C\BB$ is clearly connected for $\BB_{1,0}$ is a set with only one element, an isolated vertex.
%\bigskip

We proceed to generalize 
the coproduct to the algebraic representation. Let $\B_{\lp,\vertex}\subset\SV^{\tens \vertex}$ denote the vector space 
of all tensors representing  1VI graphs on $\lp$ loops and $\vertex$ vertices. 
Let $\B\defeq\bigoplus_{\vertex=1,\lp=0}^{\infty}\B_{\lp,\vertex}\subset\TV$. 
In 
$\B$ the analog of the coproduct $\copb$ is the map  
$\copba:\B\to\TV$ defined by the following equations: %\footnote{Note, however, that $\copba$ is not itself a coproduct on $\B$ for its range is not $\B\tens\B$.}:
\begin{eqnarray}
\copba(\one)&\defeq&\one\tens\one\,;\\
\copba(\onevi_{1,\ldots,\vertex}^\gamma)&\defeq& \frac{1}{\vertex}\sum_{i=1}^\vertex\copbai(\onevi_{1,\ldots,\vertex}^\gamma)\quad\mbox{if}\quad\vertex>1\,,\label{eq:cop1vi}
\end{eqnarray} 
where $\gamma$ denotes a 1VI graph on $\vertex$ vertices represented by $\onevi_{1,\ldots,\vertex}^\gamma\in\SV^{\tens\vertex}$. To define the maps $\copbai$ we introduce the following bijections: 
\begin{itemize}
\item
$\sigma_i:j\mapsto\left\{\begin{array}{ll} j &\mbox{if}\quad 1\leq j\leq i\\ j+1 &\mbox{if}\quad i+1\leq j\leq \vertex\\\end{array}\right.$,
\item$\nu_{i}:j\mapsto\left\{\begin{array}{ll} j &\mbox{if}\quad 1\leq j\leq i-1\\ j+1 &\mbox{if}\quad i\leq j\leq\vertex\\\end{array}\right.\,.$
\end{itemize}
In this context, for all $1\le i\le\vertex$ with $\vertex>1$, the maps $\copbai:\SV^{\tens \vertex}\to\SV^{\tens(\vertex+1)}$ are defined by the following equation:
\begin{eqnarray}
\copbai(\onevi_{1,\ldots,\vertex}^\gamma)& \defeq &\onevi_{\sigma_i(1),\ldots,\sigma_i(\vertex)}^\gamma+\onevi_{\nu_{i}(1),\ldots,\nu_{i}(\vertex)}^\gamma\\\label{eq:cop1vimain0}
&=&\onevi_{1,\ldots,\widehat{i+1},i+\two,\ldots,\vertex+1}^\gamma+\onevi_{1,\ldots,\hat{i},i+1\ldots,\vertex+1}^\gamma\,,
\label{eq:cop1vimain}
\end{eqnarray}
 where $\hat{i}$ (resp. $\widehat{i+1}$) means that the index $i$ (resp. $i+1$) is excluded from the sequence.
The tensor $\onevi^\gamma_{1,\ldots,\hat{i},\ldots,\vertex+1}$ (resp. $\onevi_{1,\ldots,\widehat{i+1},i+\two,\ldots,\vertex+1}^\gamma$) is constructed from $\onevi_{1,\ldots,\vertex}^\gamma$ by transferring  the monomial on the elementary field operators %$\phi(x)$
 which occupies the $k$th tensor factor 
to the $(k+1)$th position for all $i\le k\le\vertex$ (resp. $i+1\le k\le\vertex$).   Given a 1VI graph $\gamma$, one way to think about
the map $\copbai$ is as the operation consisting of (a) splitting the vertex $i$ in two new vertices numbered $i$ and $i+1$; (b)  transferring all the ends of edges assigned to the vertex $i$ to one of the two new vertices at a time. 
\begin{comment}
\item The counit 
$\coub:\B\to\C$ is defined by 
\begin{eqnarray}
\coub(\one)&\defeq&1\,;\\
\coub(\B_{\lp,\vertex})&:=&0\quad\mbox{if}\quad\vertex>1\,.\label{eq:coubbasic}
\end{eqnarray}
\end{itemize}
\end{comment}

Furthermore, given a bijection $\pi$ so that $i\notin\pi(\{1,\ldots,\vertex\})\subset\{1,\ldots,\vertex'\}$, 
let $\copbai(\onevi_{\pi(1),\ldots,\pi(\vertex)}^\gamma)\defeq\onevi_{\nu_i(\pi(1)),\ldots,\nu_i(\pi(\vertex))}^\gamma$. It is straightforward to verify that the maps  $\copbai$ satisfy the following %(coassociativity)
property: \begin{equation}\label{eq:coassi}\copbai\circ\copbai=\copbaiplus\circ\copbai\,,\end{equation}
where we used the same notation for 
$\copbai:\SV^{\tens \vertex}\to\SV^{\tens(\vertex+1)}$ 
on the right of either side of the above equation and $\copbai:\SV^{\tens(\vertex+1)}\to\SV^{\tens(\vertex+\two)}$ 
 on the left of the lhs. Accordingly, $\copbaiplus:\SV^{\tens(\vertex+1)}\to\SV^{\tens(\vertex+\two)}$ on the rhs. 
 
Now, 
for all $m>0$, define  the $m$th iterate of $\copbai$, $\copbai^m:\SV^{\tens\vertex}\to\SV^{\tens(\vertex+m)}$, recursively as follows:
\begin{eqnarray}\label{eq:start}
\copbai^1&\defeq&\copbai\,;\\\label{eq:iterated}
\copbai^m&\defeq&\copbai\circ\copbai^{m-1}\,,
\end{eqnarray}
where $\copbai:\SV^{\tens(\vertex+m-1)}\to\SV^{\tens(\vertex+ m)}$ in  formula (\ref{eq:iterated}).  %The last equation 
This can be written in $m$ different ways corresponding to the composition of $\copbai^{m-1}$ with each of the maps $\copbaj:\SV^{\tens(\vertex+m-1)}\to\SV^{\tens(\vertex+ m)}$ with $i\leq j\leq i+m-1$. These are all equivalent due to formula (\ref{eq:coassi}).
%%%%%%%%%%%%%%%%%%%%%%%%%%%%%%%%%%%%%%%%%%%%%%%%%%%%%%%%%%%%%%%%%%%%%%%%%%%%%%%%%%%%%%%%%%%%%%%%%%%%%%%%%%%%%%%%%%%%%%%%%%%%
\subsubsection*{Extension to connected graphs}
We now extend the maps $\copbai$ to the vector space of  all tensors representing connected graphs on at least two vertices
$\B^*\defeq\bigoplus_{k=1}^\infty\B^{\glue k}$. We proceed to define $\B^{\glue k}$. 

\bigskip

First, given two sets $A_\vertex\subset\SV^{\tens\vertex}$ and $B_{\vertex'}\subset\SV^{\tens\vertex'}$, by $A_\vertex\diamond_{i,j}B_{\vertex'}\subset\SV^{\tens(\vertex+\vertex'-1)}$; $i\in\{1,\ldots, \vertex\}, j\in\{1, \ldots,\vertex'\}$, denote the set of elements obtained by applying the map $\diamond_{i,j}$ to every ordered pair 
$(a\in A_{\vertex}, b\in B_{\vertex'})$. 

\bigskip

For all $k\ge 1$,  $\lp\ge 0$ and $\vertex\ge k+1$, define ${\B^*}^k_{\lp,\vertex}$ recursively as follows:
$${\B^*}^1_{\lp,\vertex}=\biggl\{\B_{\lp,\vertex}\biggr\}\,;$$ 
\begin{eqnarray*}
{\B^*}^\two_{\lp,\vertex}& = &\biggl\{\B_{\lp-\lp',\vertex-\vertex'}\diamond_{i,j}{\B^*}^1_{\lp',\vertex'+1} \quad \mbox{for all}\quad 
\lp'=0,\ldots,\lp, \vertex'=1,\ldots,\vertex-\two,i=1,\ldots,\vertex-\vertex', j=1,\ldots,\vertex'+1\biggr\}\,;
\end{eqnarray*}
\begin{eqnarray*}
{\B^*}^k_{\lp,\vertex} & = & \biggl\{\B_{\lp-\lp',\vertex-\vertex'}\diamond_{i,j}{\B^*}^{k-1}_{\lp',\vertex'+1} \quad \mbox{for all}\quad \lp'=0,\ldots,\lp, \vertex'=k-1,\ldots,\vertex-\two,i=1,\ldots,\vertex-\vertex', j=1,\ldots,\vertex'+1\biggr\}\,.\end{eqnarray*}
Let $${{\B'}^*}^k_{\lp,\vertex}=\bigcup_{\pi\in S_{\vertex}}\biggl\{S^\gamma_{\pi(1),\ldots,\pi(\vertex)}\vert S^\gamma_{1, \ldots, \vertex}\in {\B^*}^k_{\lp,\vertex}\biggr\}\,,$$ where  %$\sigma$ is a permutation of $\{1,\ldots,\vertex\}$
$S_\vertex$ denotes the symmetric group on the set $\{1,\ldots,\vertex\}$ and $S^\gamma_{\pi(1),\ldots,\pi(\vertex)}$ is given by formula (\ref{eq:sigma}).
Finally, for all $k\ge 1$, define $$\B^{\diamond k}\defeq\bigoplus_{\lp=0,\vertex=k+1}^\infty{{\B'}^*}^k_{\lp,\vertex}\,.$$ 
The elements of $\B^{\diamond k}$ are clearly  tensors representing connected graphs on $k$ 1VI components (i.e., maximal 1VI subgraphs \cite{handbook}). 
By equations (\ref{eq:glue}) and  (\ref{eq:sigma}), these 
may be seen as  monomials on tensors representing 1VI graphs, say,
 $\onevi_{\sigma(1),\ldots,\sigma(\vertex)}^{\gamma}$, 
with the componentwise product $\cdot:\SV^{\tens\vertex}\times\SV^{\tens\vertex}\to\SV^{\tens\vertex}$ so that repeated indices correspond to articulation vertices of the associated graphs. In this context, an arbitrary connected graph, say, $\gamma$, on $\vertex\ge\two$ vertices and $k\ge1$ 1VI components 
yields 
$$\prod_{a=1}^k\onevi_{\sigma_a(1),\ldots,\sigma_a(\vertex_a)}^{\gamma_a}\,,$$
where for all $1\le a\le k$, $\onevi_{\sigma_a(1),\ldots,\sigma_a(\vertex_a)}^{\gamma_a}\in\SV^{\tens\vertex}$, $\gamma_a$ is a  1VI graph on $\two\leq\vertex_a\leq\vertex$ vertices represented by $\onevi_{1,\ldots,\vertex_a}^{\gamma_a}\in\SV^{\tens \vertex_a}$ and 
  $\sigma_a:\{1,\ldots,\vertex_a\}\to X_a\subset\{1,\ldots,\vertex\}$ is a bijection.  

We now extend the map $\copba\defeq\frac{1}{\vertex}\sum_{i=1}^{\vertex}\copbai$ to $\B^*$ by requiring the maps $\copbai$ to satisfy the following condition:
\begin{equation}
\copbai(\prod_{a=1}^k\onevi_{\sigma_a(1),\ldots,\sigma_a(\vertex_a)}^{\gamma_a})\defeq\prod_{a=1}^k\copbai(\onevi_{\sigma_a(1),\ldots,\sigma_a(\vertex_a)}^{\gamma_a})\,.
\label{eq:algmapII}
\end{equation}
Given a connected graph $\gamma'$, 
the map $\copbai$ may be thought of as a way of (a) splitting the vertex $i$ in two new vertices numbered $i$ and $i+1$; 
(b) distributing  the 1VI components  sharing the vertex $i$ between  the two new ones in all possible ways.  Analogously, the action of the maps  $\copbai^{m}$  consists of (a) splitting the vertex $i$ in $m+1$ new vertices numbered $i,i+1,\ldots,i+m$; (b) distributing  the 1VI components  sharing the vertex $i$ between  the $m+1$ new ones in all possible ways. 
\begin{comment} Finally, the counit 
$\coub:\B^*\to\C$ is defined by 
\begin{eqnarray}
\cou(\one)&\defeq&1\,;\\
\cou(\prod_{a=1}^k\onevi_{\sigma_a(1),\ldots,\sigma_a(\vertex_a)}^{\gamma_a})&\defeq&0\quad \mbox{if}\quad k>0\,.
\end{eqnarray}
\end{comment}
%%%%%%%%%%%%%%%%%%%%%%%%%%%%%%%%%%%%%%%%%%%%%%%%%%%%%%%%%%%%%%%%%%%%%%%%%%%%%%%%%%%%%%%%%%%%%%%%%%%%%%%%%%%%%%%%%%%%%%%%%%%%%%%%%%%%%%%%%%%%%%%%%%%%%%%%%%%%%%%%%%%%%%%%%%%%%%%%%%%%%%%%%%%%%%%%%%%%%%%%%%%%%%%%%%%%%%%%%%%%%%%%%%%%%%%%%%%%%%%%%%%%%%%%%%%%%%%%%%%%%%%%%%%%%%%%%%%%%%%%%%%%%%%%%%%%%%%%%%%%%%%%%%%%%%%%%%%%%%%%%%%%%%%%%%%%%%%%%%%%%%%%%%%%%%%%%%%%%%%%%%%%%%%%%%%%
%\input{linear}
\section{Linear maps}\label{sec:linear}
We give the linear maps to be used in the following to generate 1VI and 1PI graphs. 

\bigskip

We begin by recalling the definition of the truncated coproduct given in Section VI C of Ref. \onlinecite{MeOe:npoint}. 
The
coproduct restricted to the subspace $\SVn\subset\SV$ is a map
$\SVn\to \bigoplus_{i=0}^n \SVi\tens \SVni$. Removing those components
of the direct sum where at least one of the target tensor factors is
$\SVzero$ yields the  following map, called \emph{truncated coproduct}: $$\copt:\SVn\to \bigoplus_{i=1}^{n-1} \SVi\tens
\SVni\,.$$ For
instance,
\begin{gather*}
\cop_{\ge 1}(\one)=0,\qquad \cop_{\ge 1}(\phi(x))=0,\\
\cop_{\ge 1}(\phi(x)\phi(y))=\phi(x)\tens\phi(y)
 +\phi(y)\tens\phi(x)\,.
\end{gather*}
Furthermore, we define the  maps $\Qtialg$ in analogy with the maps $\Qi$ given in Section IV of Ref. \onlinecite{MeOe:npoint}. To this end, we combine the elements $\Rop_{i,i+1}^\multiedges$ (viewed as operators on $\SV^{\tens\vertex}$ by multiplication)  with the truncated coproduct applied to the $i$th tensor factor of $\SV^{\tens\vertex}$, i.e., $\copti\defeq
\id^{\tens{i-1}}\tens\copt\tens\id^{\tens{\vertex-i}}$. That is,  for all $1\le i\le\vertex$, define
$$\Qtialg\defeq\frac{1}{\two(\multiedges-1)!}\Rop_{i,i+1}^\multiedges\circ\copti:\SV^{\tens\vertex}\to \SV^{\tens{\vertex+1}}\,.$$  
In terms of graphs, the maps $\Qtialg$  are the algebraic equivalent of the maps $\Qti$  given in  Ref. \onlinecite{Me:biconn}.  We refer the reader to that paper for the precise definition. In plain English,
the maps $\Qtialg$ produce a graph with $\vertex+1$ vertices from one with $\vertex$ vertices in the following way:
\begin{enumerate}[(a)]
\item split the vertex $i$ into two new vertices numbered $i$ and $i+1$;
\item distribute the ends of edges ending on the split vertex between the two new ones in all possible ways so that each vertex is assigned with at least one end of edges;
\item connect the two new vertices with  $\multiedges$ internal edges.
\end{enumerate}
For $\multiedges=1$, on the algebraic level, the definition of the maps $\Qtione$ (or $\Qi$ of Ref. \onlinecite{MeOe:npoint}) generalizes the application $L=(\phi\tens\phi)\cdot\Delta$ given in Section 3 of  Ref. \onlinecite{Liv:rigidity}.
On the level of graphs, this definition generalizes the fundamental operation given in Ref. \onlinecite{103} to all partitions of the set of ends of edges assigned to the vertex $i$. 
\begin{figure}[t]
\begin{center}
\includegraphics[width=5.5cm, height=1.7cm]{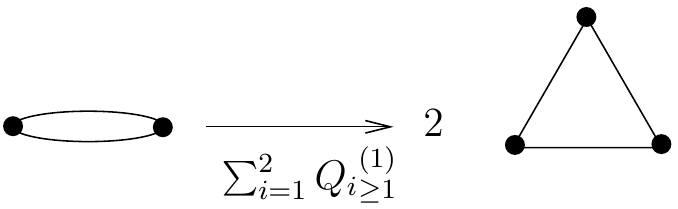}
\end{center}
\caption{The linear combination of graphs obtained by applying   $\sum_{i=1}^{\two}\Qtione$ to a cycle on two vertices.}
\label{fig:qi}
\end{figure}
For instance, to illustrate the action of the maps $\Qtione$, let $\gamma$ be a cycle on two vertices represented by $\onevi^{C_\two}_{1,\two}=\Rop_{1,\two}^\two\in\SV^{\tens\two}$. Applying  $\sum_{i=1}^{\two}\Qtione$ to $\gamma$ yields:
\begin{eqnarray}\nonumber
\sum_{i=1}^{\two}\Qtione(\onevi^{C_\two}_{1,\two})&=&\frac{1}{\two}(\Rop_{1,\two}\circ{\cop_1}_{\ge 1}+\Rop_{\two,3}\circ{\cop_\two}_{\ge 1})(\Rop_{1,\two}^\two)\\
\label{eq:exampleQ}
&=&\two\Rop_{1,\two}\cdot\Rop_{\two,3}\cdot\Rop_{1,3}\,.
\end{eqnarray}
Figure \ref{fig:qi} shows the linear combination of graphs given by equation (\ref{eq:exampleQ}). 

Furthermore, on $\B^*$ we define the maps  $\Qtifacalg$ as follows:
\begin{equation}
\Qtifacalg(\prod_{a=1}^k\onevi_{\sigma_a(1),\ldots,\sigma_a(\vertex_a)}^{\gamma_a})\defeq\frac{1}{\two(\multiedges-1)!}\Rop_{i,i+1}^\multiedges\prod_{a=1}^k\copti(\onevi_{\sigma_a(1),\ldots,\sigma_a(\vertex_a)}^{\gamma_a})\,.
\label{eq:algmapI}
\end{equation}

\bigskip
We now combine the maps $\copbai^m$ with tensors representing 1VI graphs in order to produce connected graphs.

\bigskip

Fix integers $\vertex,\vertex'\ge 1$ and $1\le i\le \vertex'$. Let $\pi_i:\{1,\ldots,\vertex\}\to\{i,i+1,\ldots,i+\vertex-1\}$ 
be any bijection. Let $\gamma\in\BB$ be a 1VI graph on $\vertex$ vertices represented by the tensor  $\onevi_{1,\ldots,\vertex}^\gamma\in\SV^{\tens\vertex}$.  
The tensors 
$\onevi_{\pi_i(1),\ldots,\pi_i(\vertex)}^\gamma\in\SV^{\vertex+\vertex'-1}$  (given by formula (\ref{eq:sigma})) may be viewed as operators acting on $\SV^{\vertex+\vertex'-1}$ by multiplication. In this context, consider the following maps given by the composition of  $\onevi_{\pi_i(1),\ldots,\pi_i(\vertex)}^\gamma$ with $\copbai^{\vertex-1}$:
$$\onevi_{\pi_i(1),\ldots,\pi_i(\vertex)}^\gamma\circ\copbai^{\vertex-1}:\SV^{\tens\vertex'}\to \SV^{\tens{\vertex+\vertex'-1}}\,.$$ 
In terms of graphs, the maps $\onevi_{\pi_i(1),\ldots,\pi_i(\vertex)}^\gamma\circ\copbai^{\vertex-1}$  are the algebraic equivalent of the maps $\ri^G$  given in Ref. \onlinecite{Me:biconn}. We refer the reader to that paper for the precise definition. In plain English,
the maps $\onevi_{\pi_i(1),\ldots,\pi_i(\vertex)}^\gamma\circ\copbai^{\vertex-1}$ produce a connected graph with $\vertex+\vertex'-1$ vertices from one with $\vertex'$ vertices in the following way:
\begin{enumerate}[(a)]
\item split the vertex $i$ into $\vertex$ new vertices numbered $i$, $i+1$,$\ldots$, $i+\vertex-1$;
\item distribute the 1VI components sharing the split vertex  between the $\vertex$ new ones in all possible ways; 
\item merge the $\vertex$ new vertices into the graph $\gamma$.
\end{enumerate}
\begin{figure}[t]
\begin{center}
\includegraphics[width=10cm]{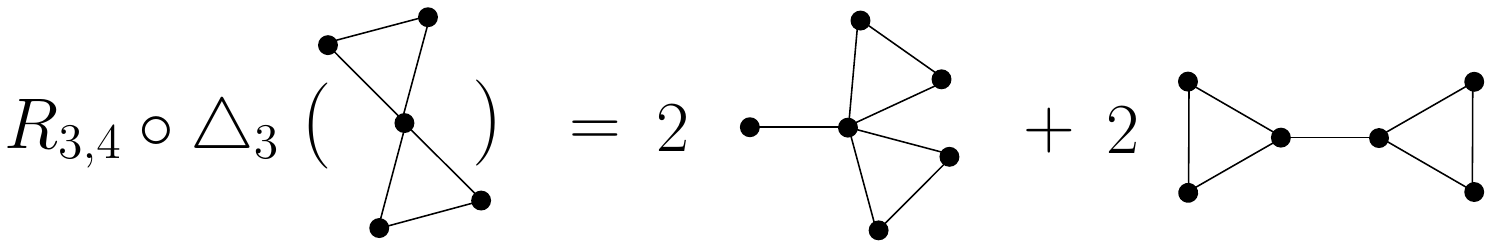}
\end{center}
\caption{The linear combination of graphs obtained by applying the map $\Rop_{3,4}\circ\copba_{3}$ to the articulation vertex of a graph consisting of two triangles sharing a vertex.}
\label{fig:rbox}
\end{figure}
To illustrate the action of the maps $\onevi_{\pi_i(1),\ldots,\pi_i(\vertex)}^\gamma\circ\copbai$, let $\gamma$ be a graph consisting of two triangles sharing a vertex. Let this be represented by $\onevi^{C_3}_{1,\two,3}\cdot\onevi^{C_3}_{3,4,5}\in\SV^{\tens5}$, where $C_3$ denotes a triangle represented by $\onevi^{C_3}_{1,\two,3}=\Rop_{1,\two}\cdot\Rop_{\two,3}\cdot\Rop_{1,3}\in\SV^{\tens3}$. Let $T_\two$ denote a $\two$-vertex tree represented  by $\onevi^{T_\two}_{1,\two}=\Rop_{1,\two}\in\SV^{\tens\two}$. Applying the map $\Rop_{3,4}\circ\copba_3$ to $\gamma$ yields:
\begin{eqnarray}\nonumber
\lefteqn{\Rop_{3,4}\circ\copba_3(\onevi^{C_3}_{1,\two,3}\cdot\onevi^{C_3}_{3,4,5})=\Rop_{3,4}\cdot\copba_3(\onevi^{C_3}_{1,\two,3})\cdot\copba_3(\onevi^{C_3}_{3,4,5})}\\\nonumber
&=&\Rop_{3,4}\cdot(\onevi^{C_3}_{1,\two,3}+\onevi^{C_3}_{1,\two,4})\cdot(\onevi^{C_3}_{3,5,6}+\onevi^{C_3}_{4,5,6})\\\label{eq:example}
&=&\Rop_{3,4}\cdot\onevi^{C_3}_{1,\two,3}\cdot\onevi^{C_3}_{3,5,6}+\Rop_{3,4}\cdot\onevi^{C_3}_{1,\two,3}\cdot\onevi^{C_3}_{4,5,6}+
 \Rop_{3,4}\cdot\onevi^{C_3}_{1,\two,4}\cdot\onevi^{C_3}_{3,5,6}+\Rop_{3,4}\cdot\onevi^{C_3}_{1,\two,4}\cdot\onevi^{C_3}_{4,5,6}\,.
\end{eqnarray}
Figure \ref{fig:rbox} shows the linear combination of graphs given by equation (\ref{eq:example}) after noticing that the first and fourth terms as well as the second and third %terms 
correspond to isomorphic graphs.
%%%%%%%%%%%%%%%%%%%%%%%%%%%%%%%%%%%%%%%%%%%%%%%%%%%%%%%%%%%%%%%%%%%%%%%%%%%%%%%%%%%%%%%%%%%%%%%%%%%%%%%%%%%%%%%%%%%%%%%%%%%%%%%%%%%%%%%%%%%%%%%%%%%%%%%%%%%%%%%%%%%%%%%%%%%%%%%%%%%%%%%%%%%%%%%%%%%%%%%%%%%%%%%%%%%%%%%%%%%%%%%%%%%%%%%%%%%%%%%%%%%%%%%%%%%%%%%%%%%%%%%%%%%%%%%%%%%%%%%%%%%%%%%%%%%%%%%%%%%%%%%%%%%%%%%%%%%%%%%%%%%%%%%%%%%%%%%%%%%%%%%%%%%%%%%%%%%%%%%%%%%%%%%%%%%%
%\input{1vi}
\section{Generating 1VI and 1PI graphs} \label{sec:gen}
We give recursion formulas to generate all 1VI and all 1PI graphs (with non-vanishing loop number) directly in the algebraic representation given in Ref. \onlinecite{MeOe:npoint}.

\subsection{1-vertex irreducible graphs}
\label{sec:1vi}
We express Theorem 4 of Ref. \onlinecite{Me:biconn}  in a completely algebraic language. 
\begin{thm}\label{thm:1vi}
For all integers $\lp\ge0$ and $\vertex> 1$,   define $\Db_{1,\ldots,\vertex}^{\lp,\vertex}\in\SV^{\tens\vertex}$ 
by the following recursion relation:
\begin{eqnarray}
\Db^{\lp,\two}_{1,\two}&\defeq&\frac{1}{\two(\lp+1)!}\Rop_{1,\two}^{\lp+1}\,;\\
\Db^{0,\vertex}_{1,\ldots,\vertex}&\defeq &0\,,\quad \vertex> \two\,; \\\label{eq:recbiconn}
\Db^{\lp,\vertex}_{1,\ldots,\vertex}  &\defeq&
\frac{1}{\lp+\vertex-1}\biggl(\sum_{\multiedges=1}^{\lp}
\sum_{i=1}^{\vertex-1}\Qtialg(\Db^{\lp+1-\multiedges,\vertex-1}_{1,\ldots,\vertex-1})+\sum_{k=2}^{\vertex-2}\sum_{\multiedges=1}^{\lp-k+1}\Qcutalg(\Daux^{\lp+1-\multiedges,\vertex-1,k}_{1,\ldots,\vertex-1})\biggr)\,,\quad \vertex> \two\,, 
\end{eqnarray}
where for all integers $k>1$, $\vertex\ge k+1$ and $\lp \ge k$, $\Daux^{\lp,\vertex,k}_{1,\ldots,\vertex}$ is given by the following recursion relation:
\begin{equation}\label{eq:aux2}
\Daux^{\lp,\vertex,\two}_{1,\ldots,\vertex}\defeq\frac{1}{\lp+\vertex-1}\sum_{\lp'=1}^{\lp-1}\sum_{\vertex'=\two}^{\vertex-1}\sum_{i=1}^{\vertex'}\sum_{j=1}^{\vertex-\vertex'+1}\biggl((\lp'+\vertex'-1)\Db^{\lp',\vertex'}_{1,\ldots,\vertex'}\glue_{i,j}\Db^{\lp-\lp',\vertex-\vertex'+1}_{1,\ldots,\vertex-\vertex'+1}\biggr)\,;
\end{equation}
\begin{equation}\label{eq:auxj}
\Daux^{\lp,\vertex,k}_{1,\ldots,\vertex}\defeq\frac{1}{\lp+\vertex-1}\sum_{\lp'=1}^{\lp-1}\sum_{\vertex'=\two}^{\vertex-1}\sum_{i=1}^{\vertex'}\biggl((\lp'+\vertex'-1)\Db^{\lp',\vertex'}_{1,\ldots,\vertex'}\glue_{i,\vertex-\vertex'+1}\Daux^{\lp-\lp',\vertex-\vertex'+1,k-1}_{1,\ldots,\vertex-\vertex'+1}\biggr)\,.
\end{equation}
Then, for fixed values of  $\vertex$ and $\lp$, 
$\Db^{\lp,\vertex}_{1,\ldots,\vertex}$ is  the weighted sum over all 1VI Feynman graphs 
with  $\lp$
loops, $\vertex$  vertices and no external edges nor self-loops,
each with weight given by the inverse of its symmetry factor.
\end{thm}
In formula (\ref{eq:recbiconn}), the 
$\Qcutalg$ summand does not appear when 
$\vertex<4$ or $\lp<\two$. 
\begin{proof}
The result follows from  Theorem 4 of Ref. \onlinecite{Me:biconn}. This is  due to the 
 correspondence between graphs and elements of $\SV^{\tens\vertex}$ and the fact that the definitions of the maps $\Qtialg$ and $\Qtifacalg$ mirror those of the maps $\Qti$ and $\Qtifac$, respectively,  given in that paper. 
\end{proof}
%%%%%%%%%%%%%%%%%%%%%%%%%%%%%%%%%%%%%%%%%%%%%%%%%%%%%%%%%%%%%%%%%%%%%%%%%%%%%%%%%%%%%%%%%%%%%%%%%%%%%%%%%%%%%%%%%%%%%%%%%%%%%%%%%%%%%%%%%%%%%%%%%%%%%%%%%%%%%%%%%%%%%%%%%%%%%%%%%%%%%%%%%%%%%%%%%%%%%%%%%%%%%%%%%%%%%%%%%%%%%%%%%%%%%%%%%%%%%%%%%%%%%%%%%%%%%%%%%%%%%%%%%%%%%%%%%%%%%%%%%%%%%%%%%%%%%%%%%%%%%%%%%%%%%%%%%%%%%%%%%%%%%%%%%%%%%%%%%%%%%%%%%%%%%%%%%%%%%%%%%%%%%%%%%%%%
%\input{1pi}
\subsection{1-particle irreducible graphs}
\label{sec:1pi}
We use Theorem \ref{thm:1vi} and the maps $\onevi_{\pi_i(1),\ldots,\pi_i(\vertex)}^\gamma\circ\copbai^{\vertex-1}$ given in Section \ref{sec:linear} to generate all 1PI graphs  following Ref. \onlinecite{Me:biconn}.

\begin{thm}\label{thm:1pi}
For all integers $\lp>0$ and $\vertex> 1$, 
define $\De^{\vertex,\lp,}_{1,\ldots,\vertex}\in\SV^{\tens\vertex}$ 
by the following recursion relation:
\begin{eqnarray}
\De^{\lp,\two}_{1,\two}&\defeq&\Db^{\lp,\two}_{1,\two}\,;\\
\label{eq:rec1pi}
\De^{\lp,\vertex}_{1,\ldots,\vertex}&\defeq&
\Db^{\lp,\vertex}_{1,\ldots,\vertex}+\frac{1}{\lp+\vertex-1}\cdot\sum_{\lp'=1}^{\lp-1}\sum_{\vertex'=\two}^{\vertex-1}\sum_{i=1}^{\vertex-\vertex'+1}
\biggl((\lp'+\vertex'-1)\Db^{\lp',\vertex'}_{\pi_i(1),\ldots,\pi_i(\vertex')}\circ\copbai^{\vertex'-1}(\De^{\lp-\lp',\vertex-\vertex'+1}_{1,\ldots,\vertex-\vertex'+1})\biggr), \vertex>\two\,.
 \end{eqnarray}
Then, for fixed values of  $\vertex$ and $\lp$, 
$\De^{\lp,\vertex}_{1,\ldots,\vertex}$ is  the weighted sum over all 1PI Feynman graphs 
with  $\lp$
loops, $\vertex$  vertices and no external edges nor self-loops, 
each with weight given by the inverse of its symmetry factor.
\end{thm}
\begin{proof}
The result follows from 
Theorem 10 of Ref. \onlinecite{Me:biconn}. This is  due to the 
 correspondence between graphs and elements of $\SV^{\tens\vertex}$ and the fact that the definition of the maps $\onevi_{\pi_i(1),\ldots,\pi_i(\vertex)}^\gamma\circ\copbai^{\vertex-1}$ mirrors that of the maps $\ri^G$ 
of that paper. 
\end{proof}

We now consider 1PI graphs with self-loops and external edges
allowed.  
Recall the elements 
$\Top_i\defeq\frac{1}{2}\Rop_{i,i}\in\SV^{\tens\vertex}$ 
with $1\le i\le\vertex$, given in  Ref. \onlinecite{MeOe:loop}. Moreover, let $\RVk$ denote the vector space generated by $T_1^k$, with $\Top_1^0=\one$.
Let
$\RV=\bigoplus_{k=0}^\infty \RVk\subset\SV$, where $\RVzero\defeq\C1$. Now, define $\delta: \RV\to\RV\tens\RV; \one\mapsto\one\tens\one; \Top_{1}^k\mapsto(\Top_1+\Top_\two)^k$.
The map $\delta$ is clearly coassociative: $(\delta\tens\id)\circ\delta=(\id\tens\delta)\circ\delta$. Moreover, for all $m\ge0$, define $\delta^m:\RV\to\RV^{\tens(m+1)}$ as follows: $\delta^0\defeq\id$ and $\delta^m\defeq(\delta\tens\id^{\tens(m-1)})\circ\delta^{m-1}$. The proposition is now established.
\begin{prop}\label{prop:selfloops}
Fix an integer $\ext\ge0$ as well as operator labels $x_1,\ldots,x_n$.   For all integers $\lp\ge1$, $\lp'\ge 0$ and $\vertex\ge1 $, define $\Ds^{\lp+\lp',\vertex}:\SV\to\SV^{\tens\vertex}$  as follows:
\begin{eqnarray}
\Ds^{\lp',1}&\defeq&\frac{1}{\lp'!}\Top_1^{\lp'}\circ\id\,;\\
\Ds^{\lp+\lp',\vertex}&\defeq&\frac{1}{\lp'!}(\De^{\lp,\vertex}_{1,\ldots,\vertex}\cdot\delta^{\vertex-1}(\Top_1^{\lp'}))\circ\cop^{\vertex-1}\,,\vertex\ge 2\,.\label{eq:selfloops}
\end{eqnarray}
Then, 
$\Ds^{\lp+\lp',\vertex}(\phi(x_1)\cdots\phi(x_n))$ is the weighted sum over all 1PI Feynman graphs with $\lp$ loops (excluding self-loops),
$\lp'$ self-loops, $\vertex$ vertices and $n$ external edges whose end points are  labeled 
$x_1,\dots,x_n$, each with weight given by the inverse of its symmetry factor.
\end{prop}
\begin{proof} 
Clearly, 
$\delta^{\vertex-1}(\Top_1^{\lp'})=\sum_{{\lp'}_1,\ldots,{\lp'}_\vertex}\binom{\lp'}{{\lp'}_1,\ldots,{\lp'}_\vertex}\Top_1^{{\lp'}_1}\cdot\ldots\cdot\Top_{\vertex}^{{\lp'}_\vertex}$, where the sum runs over all partitions of ${\lp'}$ into $\vertex$ non-negative integers. That is,  for all $i=1,\ldots, \vertex$, $\lp'_i\ge0$ with $\sum_i^\vertex{\lp'}_i=\lp'$. Therefore, the result follows straightforwardly from the symmetry property of the multinomial coefficients: $\binom{\lp'}{{\lp'}_1,\ldots,{\lp'}_{\vertex}}=\binom{\lp'}{{\lp'}_{\sigma(1)},\ldots,{\lp'}_{\sigma(\vertex)}}$, where $\sigma$ denotes any permutation of $\{1,\ldots,\vertex\}$.
\end{proof}
%%%%%%%%%%%%%%%%%%%%%%%%%%%%%%%%%%%%%%%%%%%%%%%%%%%%%%%%%%%%%%%%%%%%%%%%%%%%%%%%%%%%%%%%%%%%%%%%%%%%%%%%%%%%%%%%%%%%%%%%%%%%%%%%%%%%%%%%%%%%%%%%%%%%%%%%%%%%%%%%%%%%%%%%%%%%%%%%%%%%%%%%%%%%%%%%%%%%%%%%%%%%%%%%%%%%%%%%%%%%%%%%%%%%%%%%%%%%%%%%%%%%%%%%%%%%%%%%%%%%%%%%%%%%%%%%%%%%%%%%%%%%%%%%%%%%%%%%%%%%%%%%%%%%%%%%%%%%%%%%%%%%%%%%%%%%%%%%%%%%%%%%%%%%%%%%%%%%%%%%%%%%%%%%%%%%
%\input{npoint}
\subsection{1PI $n$-point functions} \label{sec:npoint}
We turn to the interpretation of $\Ds^{\lp+\lp',\vertex}$ given by equation (\ref{eq:selfloops}) in terms of Feynman graphs
and $n$-point functions.
We emphasize that the discussion here applies only to bare $n$-point functions  for renormalization is outside the scope of the present work. Moreover, the results are valid for any quantum field  theory. 

\bigskip

In all that follows, let $\lp$ and $\lp'$  denote the loop number (excluding self-loops) and the number of self-loops, respectively,  of any Feynman graph.
Now, by $\fctpi$, denote the 1PI vertex functions, i.e.,  the interaction vertices of each of the terms in  the perturbative expansion of the 1PI $n$-point  functions. By $\opi^{\lp+\lp',\vertex}$, denote the $\lp+\lp'$-loop and $\vertex$-vertex
contribution to the ensemble $\opi$  of 1PI  $n$-point functions. The $\lp+\lp'$-loop order
contribution $\opi^{\lp+\lp'}$ to $\opi$ and $\opi$ itself, are given by
\begin{equation*}
\opi^{\lp+\lp'}=\sum_{\vertex=1}^\infty \opi^{\lp+\lp',\vertex},\qquad
\opi=\sum_{\lp+\lp'=0}^\infty \opi^{\lp+\lp'} .
\end{equation*}
 All  vertex order contributions are captured by
the following corollary.

\begin{cor}
\label{cor:loopexpansion}
For $v\ge 1$:
\[
 \opi^{\lp+\lp',\vertex}=\fctpi^{\tens \vertex}\circ\Ds^{\lp+\lp',\vertex} .
\]
\end{cor}
%%%%%%%%%%%%%%%%%%%%%%%%%%%%%%%%%%%%%%%%%%%%%%%%%%%%%%%%%%%%%%%%%%%%%%%%%%%%%%%%%%%%%%%%%%%%%%%%%%%%%%%%%%%%%%%%%%%%%%%%%%%%%%%%%%%%%%%%%%%%%%%%%%%%%%%%%%%%%%%%%%%%%%%%%%%%%%%%%%%%%%%%%%%%%%%%%%%%%%%%%%%%%%%%%%%%%%%%%%%%%%%%%%%%%%%%%%%%%%%%%%%%%%%%%%%%%%%%%%%%%%%%%%%%%%%%%%%%%%%%%%%%%%%%%%%%%%%%%%%%%%%%%%%%%%%%%%%%%%%%%%%%%%%%%%%%%%%%%%%%%%%%%%%%%%%%%%%%%%%%%%%%%%%%%%%%
%\input{acknowledge}
\begin{acknowledgments}
The author would like to thank Christian Brouder  for his
careful reading and helpful comments on the manuscript. 
The research was supported through the fellowship SFRH/BPD/48223/2008 provided by the Portuguese Science and Technology Foundation.
\end{acknowledgments}
%%%%%%%%%%%%%%%%%%%%%%%%%%%%%%%%%%%%%%%%%%%%%%%%%%%%%%%%%%%%%%%%%%%%%%%%%%%%%%%%%%%%%%%%%%%%%%%%%%%%%%%%%%%%%%%%%%%%%%%%%%%%%%%%%%%%%%%%%%%%%%%%%%%%%%%%%%%%%%%%%%%%%%%%%%%%%%%%%%%%%%%%%%%%%%%%%%%%%%%%%%%%%%%%%%%%%%%%%%%%%%%%%%%%%%%%%%%%%%%%%%%%%%%%%%%%%%%%%%%%%%%%%%%%%%%%%%%%%%%%%%%%%%%%%%%%%%%%%%%%%%%%%%%%%%%%%%%%%%%%%%%%%%%%%%%%%%%%%%%%%%%%%%%%%%%%%%%%%%%%%%%%%%%%%%%%

%\bibliographystyle{abbrv}
%\bibliography{bib1001b}

%%%%%%%%%%%%%%%%%%%%%%%%%%%%%%%%%%%%%%%%%%%%%%%%%%%%%%%%%%%%%%%%%%%%%%%%%%%%%%%%%%%%%%%%%%%%%%%%%%%%%%%%%%%%%%%%%%%%%%%%%%%%%%%%%%%%%%%%%%%%%%%%%%%%%%%%%%%%%%%%%%%%%%%%%%%%%%%%%%%%%%%%%%%%%%%%%%%%%%%%%%%%%%%%%%%%%%%%%%%%%%%%%%%%%%%%%%%%%%%%%%%%%%%%%%%%%%%%%%%%%%%%%%%%%%%%%%%%%%%%%%%%%%%%%%%%%%%%%%%%%%%%%%%%%%%%%%%%%%%%%%%%%%%%%%%%%%%%%%%%%%%%%%%%%%%%%%%%%%%%%%%%%%%%%%%%
\end{document}